\documentclass[aps,preprint,showpacs,floats,epsf,epsfig,nofootinbib,12pt]{revtex4}
\usepackage{graphicx}
\usepackage{epsfig}
\usepackage[dvips]{color}

\def\be{\begin{eqnarray}}
\def\en{\end{eqnarray}}

\begin{document}

\title{Determination of the mixing between active neutrinos and sterile neutrino through the quark-lepton complementarity and self-complementarity }

\vspace{1cm}

\author{ Hong-Wei Ke$^{1}$   \footnote{khw020056@hotmail.com}, Tan Liu$^{1}$ and
        Xue-Qian Li$^2$\footnote{lixq@nankai.edu.cn}  }

\affiliation{  $^{1}$ School of Science, Tianjin University, Tianjin 300072, China \\
  $^{2}$ School of Physics, Nankai University, Tianjin 300071, China }

\vspace{12cm}

\begin{abstract}
It is suggested that there is an underlying symmetry which relates
the quark and lepton sectors. Namely, among the mixing matrix
elements of CKM for quarks and PMNS for leptons there exist
complementarity relations at a high energy scale (such as the
see-saw or even the GUT scales). We Assume that the relations
would remain during the matrix elements running down to the
electroweak scale. Observable breaking of the rational relation is
attributed to existence of sterile neutrinos which mix with the
active neutrino to result in the observable PMNS matrix. We show
that involvement of a sterile in the (3+1) model, induces that
$|U_{e4}|^2= 0.040$, $|U_{\mu4}|^2= 0.009$ and $\sin^2
2\alpha=0.067$. We also find a new self-complementarity
$\vartheta_{12}+\vartheta_{23}+\vartheta_{13}+\alpha\approx90^\circ$.
The numbers are generally consistent with those obtained by
fitting recent measurements, especially in this scenario, the
existence of a sterile neutrino  does not upset the LEP data i.e.
the number of neutrino types is very close to 3.


\end{abstract}

\maketitle

\section{Introduction}

By the recent observation, the neutrino masses are much lighter
than the corresponding leptons, but the origin of the neutrino
masses remains as a mystery. Even so, in analog to the quark
sector, the neutrino flavor eigenstates are different from their
mass eigenstates, so the Pontecorvo-Maki-Nakawaga-Sakata (PMNS)
matrix appears \cite{Maki:1962mu,Pontecorvo:1967fh}. On the
aspect, the quark sector has been treated separately and a
corresponding mixing matrix Cabibbo- Kobayashi-Maskawa (CKM)
\cite{Cabibbo:1963yz,Kobayashi:1973fv,PDG12} plays the role. It is
well known that to cancel the gauge anomaly, quark and lepton
sectors must exist simultaneously and have the same generations.
Therefore, one is tempted by the one-to-one correspondence of
quark and lepton to conjecture there might be some intimate
relations between the two sectors. Even though at the practical
world, the two sectors look quite differently, one may consider
that at very high scales such as the see-saw, Grand Unification or
even Planck scales, they originate from the same or at least
related sources. Therefore there might exist a large symmetry
which relates the two sectors. {Indeed, at the high energy scale
of about $10^{15}$ GeV, where the strong, electromagnetic (EM) and
weak interactions are unified into a large symmetry such as SU(5),
SO(10), E6 etc. \cite{Langacker:1980js}, quarks and leptons may
reside in the same representations of the large group, therefore
by the Grand Unification Theories (GUTs) it is natural to expect
that such complementary relations may exist. Earlier in 2004,
Raidal \cite{Raidal:2004iw} suggested that GUTs may relate the
quark and lepton sectors and predicted some phenomenological
consequences, and  Ma $et\, al.$\cite{Zhang:2012zh} considered it
as the theoretical base of complementarity.  }

If the two sectors indeed originate from a large symmetry, even though during the process of running down from
higher energy scale to our practical electro-weak scale many quantities look different, some of the relations may
remain.

With a certain parametrization the quark-lepton
complementarity\cite{Minakata:2004xt} and
self-complementarity\cite{Altarelli:2009kr,Zheng:2011uz,Zhang:2012pv,Haba:2012qx}
are noticed. Indeed such relations are approximate. Motivated by
the picture described above, we assume that the complementarity
and self-complementarity are exact and guaranteed by the residual
symmetry which even though is not clear yet. On other aspect, the
experimental measurements show that such relations are only
approximate. One may think that such deviations are due to
measurement errors, or there exists new physics whose existence
results in the declination from exact complementarity and
self-complementarity. The goal of this work is to search for a
possible new physics scenario which may cause such a declination.

The short-baseline neutrino oscillations indicates there may exist
light sterile neutrinos if CPT-invariance is
conserved\cite{Sorel:2003hf,Maltoni:2007zf,Karagiorgi:2009nb}. The
sterile neutrinos do not directly participate in weak interaction,
but may mix with the active neutrinos of three generations.
Therefore they would make substantial contributions to the
observable physical quantities via the mixing.

The mixing among active neutrinos and sterile neutrinos produces
an extended $4\times 4$ PMNS matrix
\cite{Girardi:2014wea,Kisslinger:2013sba} with certain parameters.
In the previous $3\times 3$ PMNS matrix
\cite{Zheng:2011uz,Zhang:2012pv,Zhang:2012zh} the mixing with
sterile neutrinos was not included, so that the quark-lepton
complementarity and self-complementarity are approximate. Now we
assume the quark-lepton complementarity and self-complementarity
to be exact, whereas the mixing among active and sterile neutrinos
causes the apparent declination. The starting point of this work:
the mixing angles for quarks and leptons possess an exact
complementarity, but contaminated by existence of sterile
neutrinos.

In this paper we will employ the scenario with three active neutrinos plus
one sterile neutrino (3+1). Thus we may fix mixing angles between the light
sterile neutrino with the active ones in term of the assumed quark-lepton complementarity and
self-complementarity. Though the scenario is simple, its prediction generally coincides with the
experimental observation, thus the present data do not suggest us to
abandon the simple version\cite{Giunti:2013aea}.

To be explicit, we re-state our strategy as:   By supposing  the quark-lepton
complementarity and self-complementarity to be exact, we
calculate the mixing matrix elements
$|U_{e1}|$, $|U_{e2}|$, $|U_{e3}|$, $|U_{\mu3}|$ and $|U_{\tau3}|$
of the original matrix PMNS (i.e. without mixing with the sterile neutrino). Then the mixing matrix is extended to
a $4\times 4 $ matrix which includes mixing between
active neutrinos and a sterile neutrino as suggested in literature, and the  3$\times$3 sub-matrix at the left-upper corner
of the 4$\times $4 matrix is the practical PMNS matrix.
Comparing the matrix elements with the data one can fix the
mixing angles between the sterile neutrino and the active neutrinos, and
determine the weak CP phase. (see the later context for details).

Finally, we test the scenario by calculating the number neutrino generations, which is determined by the
LEP data very accurately as very close to 3. Our result shows that this number is perfectly respected in the new scenario.

The paper is organized as follows. After the introduction we describe
our detailed strategy  and derivation of relevant formulas in section II. In section III, we
present our numerical results along with all the inputs and discuss both experimental and theoretical errors.
In section IV we will make a summary.

\section{the mixing of fermions and quark-lepton complementarity and self-complementarity}
In this section we show explicitly how to fix the mixing angles between the
active neutrinos and sterile neutrino and the CP-phase
under the hypothesis of the exact quark-lepton complementarity and
self-complementarity.

\subsection{the mixing of fermions in SM }

The mixing among quarks or leptons is described by the CKM and PMNS matrices which appear in the
weak charged currents. The quark sector involves the u-type and d-type quarks, whereas the leptonic sector
involves  neutrinos and charged leptons. The relevant Lagrangian is
\begin{eqnarray} \label{Lag}
\mathcal{L}=\frac{g}{\sqrt{2}}\bar {U}_L\gamma^\mu V_{CKM} D_L
W^+_{\mu}-\frac{g}{\sqrt{2}}\bar E_L \gamma^\mu V_{PMNS} N_L
W^+_{\mu}+h.c.,
\end{eqnarray}
where $U_L=(u_L, c_L, t_L)^T$,  $D_L=(d_L, s_L, b_L)^T$,
$E_L=(e_L, \mu_L, \tau_L)^T$and  $N_L=(\nu_1, \nu_2, \nu_3)^T$.
$V_{CKM}$ and $V_{PMNS}$ are the CKM  and PMNS matrices
respectively. If there were no sterile neutrino, both quark and lepton sectors
contain three generations, so their mixing matrices are similar.
As is well known that real physics is independent of any parametrization schemes, so
it is convenient to set $V_{CMS}$ and $V_{PMNS}$ in the P1
parametrization\cite{Zhang:2012zh} as

    \begin{equation}\label{M1}
      V=\left(\begin{array}{ccc}
        U_{e1} &U_{e2} &U_{e3} \\
         U_{\mu1} &   U_{\mu2} &  U_{\mu3}\\
          U_{\nu1} & U_{\nu2} & U_{\nu3}
      \end{array}\right)=\left(\begin{array}{ccc}
        c_{12}c_{13} & s_{12}s_{13} & s_{13}\\
       -c_{12}s_{23}s_{13} -s_{12}c_{23}e^{i \delta(\delta')} & -s_{12}s_{23}s_{13}+c_{12}c_{23}e^{i\delta(\delta')} & s_{23}c_{13}\\
      -c_{12}s_{23}s_{13} + s_{12}s_{23}e^{i\delta(\delta')} & -s_{12}s_{23}s_{13}-c_{12}s_{23}e^{i\delta(\delta')} & c_{23}c_{13}
      \end{array}\right).
  \end{equation}
Here $s_{ij}$ and $c_{ij}$ denote $\sin\theta_{ij}(\sin\vartheta_{ij})$ and
$\cos\theta_{ij}(\cos\vartheta_{ij})$ with $i,j=1,2,3$. In this work, we use $\theta_{ij}$ for quark sector and
$\vartheta_{ij}$ for lepton sector respectively.

Thanks to hard experimental measurements on the weak processes
where the CKM matrix is involved, the mixing parameters for the quark sector are more precisely fixed and their
central values\cite{Zhang:2012pv} are
\begin{eqnarray} \label{qmix1}
\theta_{12}=13.023^\circ, \theta_{23}=2.360^\circ,
\theta_{13}=0.201^\circ, \delta=69.10^\circ.
\end{eqnarray}
Definitely, certain experimental errors still exist and they would cause theoretical uncertainties in our predictions
on the PMNS parameters. We will discuss that issue later.

The parameters in the $3\times 3$ PMNS matrix which are determined by the measured data
\cite{Zhang:2012pv} are
\begin{eqnarray} \label{qmix2}
\vartheta_{12}=33.65^\circ, \vartheta_{23}=38.41^\circ,
\vartheta_{13}=8.93^\circ,
\end{eqnarray}
which are directly measured by the neutrino-involved experiments, especially
the neutrino oscillations.

As we discussed above, among the CKM and PMNS matrix elements, there are
complementarity and self-complementarity relations. In the P1 parametrization,
the relations reduce to some direct relations among the mixing angles.
The quark-lepton complementarity suggests
$\theta_{12}+\vartheta_{12}\approx45^\circ$,
$\theta_{23}+\vartheta_{23}\approx45^\circ$ and the
self-complementarity requires
$\vartheta_{12}+\vartheta_{13}\approx\vartheta_{23}$ to be held.

Comparing with data, one immediately notices that even though
those relations are in a good approximation, obvious deviation of the obtained mixing matrix from the data,
 \begin{equation}\label{M2}
      V'=\left(\begin{array}{ccc}
        U_{e1}' &U_{e2}' &U_{e3}' \\
         U_{\mu1}' &   U_{\mu2}' &  U_{\mu3}'\\
          U_{\nu1}' & U_{\nu2}' & U_{\nu3}'
      \end{array}\right),
  \end{equation}
demands an explanation.
That deviation happens in the
scenario with only three types of neutrinos as required by the
standard model, so when the theory is extended to involve new
components, the problem would be easily solved.

\subsection{the mixing of neutrinos  beyond SM
}

In some previous works, the authors introduced one or more sterile
neutrinos to explain the data of short-baseline neutrino
oscillation\cite{Sorel:2003hf,Maltoni:2007zf,Karagiorgi:2009nb}.
In this work we consider the model of three active neutrinos
mixing with one sterile neutrino ($\nu_s$). The sterile neutrino does
not directly participate the weak interaction,
so before taking into account its mixing with active neutrinos,
the weak interaction Lagrangian for
leptonic sector is
\begin{equation}\label{ab}-\frac{g}{\sqrt{2}} \left(\begin{array}{ccc}
        \bar e_L & \bar\mu_L& \bar\tau_L
      \end{array}\right)\gamma^{\mu}\left(\begin{array}{cccc}
        U_{e1}' &U_{e2}' &U_{e3}'&0 \\
         U_{\mu1}' &   U_{\mu2}' &  U_{\mu3}'&  0 \\
          U_{\tau1}' & U_{\tau2}' & U_{\tau3}'& 0
      \end{array}\right)\left(\begin{array}{c}
       \nu_1 \\
         \nu_2\\
          \nu_3\\
          \nu_s \end{array}\right)_LW^+_{\mu}+h.c. \,.
    \end{equation}
Apparently as the active neutrinos mix with the sterile neutrino,
the values of the mixing matrix elements in eqs.(\ref{M2}) and
(\ref{ab})  are definitely affected. Once appropriate mixing
parameters are chosen, these modified mixing matrix elements may
coincide with the available data. By contrary, if one cannot fix a
set of such mixing parameters to make the new matrix elements to
meet the data, the model would fail. Later we will show that the
adopted scenario succeeds, i.e. the newly obtained PMNS matrix elements
are generally consistent with the data and the theoretical uncertainties
are smaller than the experimental errors.


To account for the possible mixing between the
sterile neutrino and the active ones, we introduce a $4\times 4$
matrix. In a complete picture, the mixing of neutrinos (3 active
neutrinos+1 sterile neutrino) could be
 \begin{equation}\label{M3}
\left(\begin{array}{c}
       \nu_e \\
         \nu_\mu\\
          \nu_\tau\\
          \nu_s \end{array}\right)
=V_{4\times 4}\left(\begin{array}{c}
       \nu_1 \\
         \nu_2\\
          \nu_3\\
          \nu_4 \end{array}\right),
    \end{equation}
where     $\nu_s$ is the sterile neutrino   and the
extended $4\times 4$ matrix is written as
\begin{equation}\label{M4}
V_{4\times 4}=\left(\begin{array}{cccc}
        U''_{e1} &U''_{e2} &U''_{e3}&U''_{e4} \\
         U''_{\mu1} &   U''_{\mu2} &  U''_{\mu3}&  U''_{\mu4} \\
          U''_{\tau1} & U''_{\tau2} & U''_{\tau3}& U''_{\tau4}\\
           U''_{s1}& U''_{s2}& U''_{s3}& U''_{s4}
      \end{array}\right),
    \end{equation}
which can be realized in a rotation
\cite{Girardi:2014wea,Kisslinger:2013sba}
 \begin{equation}\label{u1}
V_{4\times 4}=R_{23}\phi R_{13}R_{12}R_{14}R_{24}R_{34},
    \end{equation}
and the relevant matrices $R_{23}, R_{13}, R_{12}, R_{14}, R_{24},
R_{34}$ and $\phi$ are simple and straightforward  as done in
literature, however for readers' convenience we present them in
the appendix. It is noted that the left upper $3\times 3$
sub-matrix corresponds to the measured PMNS mixing matrix whose
elements are fixed by the neutrino oscillation experiments.

\subsection{the strategy to fix the mixing parameters }
As a sterile neutrino is introduced into the model, there would be more free
parameters.
We have obtained the modules of $U_{e1}'$, $U_{e2}'$,
$U_{e3}'$, $U_{\mu3}'$ and $U_{\nu3}'$ in eq.(\ref{M2}).
Taking into account the mixing with the sterile neutrino,  the
elements $U_{e1}'$, $U_{e2}'$, $U_{e3}'$, $U_{\mu3}'$ and
$U_{\nu3}'$ are modified to $U''_{e1}$, $U''_{e2}$, $U''_{e3}$,
$U''_{\mu3}$ and $U''_{\nu3}$ in eq.(\ref{M4}). By adjusting the
mixing parameter, we can make those elements to eventually
coincide with the measured values.

Supposing that the quark-lepton
complementarity and self-complementarity hold for the 3-generation neutrino
structure the central values of the measured CKM
matrix elements for quarks would fully determine
$\theta_{12}=(13.023\pm0.038)^\circ,
\theta_{23}=(2.360\pm0.052)^\circ$ and $
\theta_{13}=(0.201\pm0.009)^\circ $, then we can obtain
$\vartheta'_{12}=(31.977\pm0.038)^\circ$,
$\vartheta'_{23}=(42.640\pm 0.052)^\circ$,
$\vartheta'_{13}=(10.663\pm 0.014)^\circ$
which are deviate from the values given in Eq.(4). Let us re-write
the PMNS matrix in terms of the obtained angles as
 \begin{equation}\label{M5}
      |V'|=\left(\begin{array}{ccc}
       0.834{\pm 0.001} &0.520{\pm0.001} & 0.185\pm0.001\\
        - & -&0.666{\pm0.001}\\
       - &
      -
        &0.723{\pm0.001}
      \end{array}\right).
  \end{equation}
The corresponding experimental values in the $3\times 3$
$V_{PMNS}$ is\cite{Zhang:2012pv}

    \begin{equation}\label{M8}
      |V_{PMNS}|=\left(\begin{array}{ccc}
       0.822{\pm 0.011} &0.547{\pm0.016} & 0.155\pm0.008\\
        - & -&0.614{\pm0.018}\\
       - &
      -
        &0.774{\pm0.014}
      \end{array}\right).
  \end{equation}
One can notice the deviation.

Then we introduce the mixing with the sterile neutrino and
re-calculate the modules of $U''_{e1}$, $U''_{e2}$, $U''_{e3}$,
$U''_{\mu3}$ and $U''_{\tau3}$ in the $V_{4\times 4}$ matrix. Now
the numbers can be compared with the measured values of the $
V_{PMNS}$ elements. Here let us explicitly show the expression of
$|U_{e1}|$ as an example
\begin{equation}\label{M7}
      |U''_{e1}|=\cos\vartheta'_{12}\cos\vartheta'_{13}
      \cos\alpha=0.834\cos\alpha.
  \end{equation}

Comparing with the data,
\begin{equation}\label{M7}
      |U''_{e1}|=|U_{e1}|,
  \end{equation}
we fix the mixing parameters. The  other elements and CP phase
$\delta'$ are simultaneously fixed, when the $\chi^2$ methods is
employed\cite{Chiang:2004nm,Ke:2007ih}.

At last, using these parameters we complete the generalized and practical $4\times 4$
matrix and its left-upper $3\times 3$ sub-matrix is just the  practical matrix $|V_{PMNS}|$.

\section{numerical results}
There are two possible schemes for the 3+1 mixing.

1. The first scheme: the sterile neutrino mixes with the three active neutrino by different
mixing parameters, namely the there are three free parameters $\alpha$, $\beta$ and $\gamma$.

To fit the data, we set the values:
$\alpha=(0.00\pm0.02)^\circ$, $\beta=(14.19\pm0.18)^\circ$,
$\gamma=(12.46\pm 0.19)^\circ$ and CP phase
$\delta'=(0.00\pm0.01)^\circ$. The module of the PMNS matrix reads
  \begin{equation}\label{M9}
|V_{4\times 4}|=\left(\begin{array}{cccc}
       0.834\pm0.001 &0.505\pm0.001 &0.153\pm0.002&0.165\pm0.002 \\
        0.496\pm0.001 &  0.541\pm0.001 &  0.621\pm0.002&  0.277\pm0.003 \\
         0.243\pm0.001 & 0.627\pm0.001 & 0.740\pm0.001& 0.001\pm0.004\\
          0\pm0.001&0.245\pm0.004& 0.209\pm0.004& 0.947\pm0.002
      \end{array}\right).
    \end{equation}
The resultant $|U''_{e3}|$, $|U''_{\mu3}|$, and $|U''_{\tau3}|$
are close to data. Based on our calculations we have
$|U''_{e4}|^2= 0.027\pm0.004$, $|U''_{\mu4}|^2= 0.077\pm0.006$ and
sin$^2 2\alpha=0\pm0.002$. In the earlier
works\cite{Giunti:2013aea,Sorel:2003hf,Giunti:2011cp,Kopp:2011qd}
the authors carried out an  analysis of short-baseline neutrino
oscillations in the 3+1 neutrino mixing scenario. Their results
are presented in table I.

\begin{table}
\caption{ the values of $|U''_{e4}|^2$ and
 $|U''_{\mu4}|^2$ in this work and in references.} \label{tab:expect}
\begin{tabular}{c|c|c|c|c|c|c}\hline\hline
 ~~~~~~~~   &  Ref.\cite{Giunti:2013aea}  &
 ~~~Ref.\cite{Sorel:2003hf}~~~ &  ~~~Ref.\cite{Giunti:2011cp}~~~&  ~~~Ref.\cite{Kopp:2011qd}~~~ & First Scheme&Second Scheme  \\\hline
 $|U''_{e4}|^2$    &$0.03\sim0.033$    &  0.0185      & $0.027\sim0.036$ &0.0228 &$0.027\pm0.004$ &$0.040\pm0.004$   \\
  $|U''_{\mu4}|^2$    &$0.0073\sim0.014$    &  0.042      &  $0.0084\sim0.021$&- & $0.077\pm0.006$&$0.009\pm0.002$ \\
\hline\hline
\end{tabular}
\end{table}

2. The second scheme: That is a simplified version of the first
scheme,  we let  $\alpha=\beta=\gamma$ as discussed in
Ref.\cite{Kisslinger:2013sba}. And then we carry out the same
process to determine the single parameter $\alpha$. The parameters
$\alpha=(7.51\pm0.04)^\circ$ and $\delta'=(0.00\pm0.01)^\circ$ are
obtained. The modulus of corresponding PMNS matrix is
  \begin{equation}\label{M10}
|V_{4\times 4}|=\left(\begin{array}{cccc}
        0.826\pm0.001 &0.502\pm0.001 &0.161\pm0.001&0.199\pm0.002 \\
        0.492\pm0.001 &   0.561\pm0.001 &  0.659\pm0.001& 0.096\pm0.001\\
          0.241\pm0.001 & 0.645\pm0.001 & 0.724\pm0.001&0.042\pm0.001\\
          0.131\pm0.001& 0.130\pm0.001& 0.128\pm0.001& 0.974\pm0.001
      \end{array}\right).
    \end{equation}

In this scenario, which assumes the mixing  between the sterile neutrino and
the different active neutrinos is nondistinctive. Our estimates are presented in table
I.

Moreover, we find a new self-complementarity
$\vartheta'_{12}+\vartheta'_{23}+\vartheta'_{13}+\alpha\approx90^\circ$
which is a bit different from that self-complementarity relation
given in Ref.\cite{Haba:2012qx}.

As a test one would calculate the neutrino flavor number which is determined to be
3 by the LEP data. Ignoring the neutrino masses, the neutrino number is
\begin{eqnarray}
N_{\nu} &=& \sum_{\rho,\sigma=1}^4\Gamma(Z\to \bar\nu_{\rho}\nu_{\sigma})/\Gamma(Z\to \bar\nu\nu)\nonumber\\
&=& \sum_{\rho,\sigma=1}^4|\sum_{i=1}^3(V^\dagger)_{\rho
i}V_{i\sigma} |^2,
\end{eqnarray}
where $V_{i\sigma}$ is the generalized PMNS matrix which is a
$3\times 4$ matrix and not unitary. Our numerical result shows
that in this scenario, $N_{\nu}$ is 3, which is fully consistent
with the LEP measurement within a reasonable error tolerance. The
denominator of the above equation $\Gamma(Z\to \bar \nu\nu)$
stands for the partial decay width of $Z$ boson into a neutrino
pair calculated in the SM.

\section{Summary}
In this work we adopt the two
quark-lepton complementarity relations and a self-complementarity relation proposed in literatures \cite{Minakata:2004xt,Raidal:2004iw,Zheng:2011uz}
which is supposed to originate from a higher symmetry and maintain when energy scale runs down to
the electroweak scale.

Then the deviation of the determined values from the measured PMNS
matrix elements is attributed to the involvement of a sterile
neutrino. The mixing of the sterile neutrino with the active ones
results in the practical values of the PMNS matrix. Comparing with
data, we are able to determine the mixing parameters.

In this work, we choose two schemes, in the first scheme, the sterile neutrino mixes with
three different active neutrino by different parameters (i.e. $\alpha$, $\beta$ and $\gamma$ are
independent parameters which are determined by fitting data; whereas in the second scheme, we let
$\alpha=\beta=\gamma$, so that there is only one parameter to describe the mixing. The numerical
values are listed in Tab.1.

It is noted that the previous estimates on the mixing between sterile neutrino and the active ones were
obtained by fitting the data, instead, by our strategy, we start with the theoretical assumption: the
complementarity and self-complementarity. The relevant mixing elements obtained in previous literatures are quite
disperse and the only common point is that the sterile-active mixing is small, no matter how to obtain them.

By the first scheme, our prediction on $|U''_{e4}|^2$ is generally
consistent with the results given by the authors of Ref.[20,22]
(see table I), but the value of $|U''_{\mu4}|^2$ is slightly
bigger. The compatibility of reactor antineutrino anomaly was
discussed in Ref.\cite{Mention:2011rk} and the mixing parameter
sin$^2 2\alpha=0.14\pm 0.08$ was fixed when $\Delta
m_{41}^2>1.5$eV$^2$. Our estimation on $|U''_{e 4}|^2$ is
consistent also with it within a $2\sigma$ range.

For the second scheme, the numbers look differently, but the trend and consistency degree
with those given in literatures are all within the present experimental error tolerance.

The
theoretical uncertainties of our predictions originate from the measurement errors of the CKM matrix
elements which are relatively small thanks to many years of hard work.
On the contrary the experimental errors for measuring the PMNS matrix elements are
larger.
Thus, our predictions on the mixing between sterile and active
neutrinos and that obtained by others are still consistent with each others
within 1-2 $\sigma$ ranges.

Recently the Daya Bay collaboration reports their new
data\cite{An:2014mra} on the mixing between the sterile neutrino
and active neutrinos, but the errors are still too large to make a conclusive
judgement on the validity of our theory yet. The future
improved measurement may further narrow down the data ranges, so
that we can testify any theoretical ansatz and get a better
understanding on neutrinos.

\section*{Acknowledgement}

This work is supported by the National Natural Science Foundation
of China (NNSFC) under the contract No. 11375128 and 11135009. We
greatly benefit from Prof. Lam's lecture given in Nankai
University several years ago.

\appendix

\section{}
 \begin{equation}
R_{23}=\left(\begin{array}{cccc}
       1 &0 &0&0 \\
        0 &  C_{23} &  S_{23}& 0 \\
        0 & -S_{23} & C_{23}& 0\\
          0& 0& 0&1
      \end{array}\right),
      R_{13}=\left(\begin{array}{cccc}
       C_{13} &0 &S_{13}&0 \\
        0 & 1 &  0& 0 \\
       -S_{13} & 0 & C_{13}& 0\\
          0& 0& 0&1
      \end{array}\right),
 R_{12}=\left(\begin{array}{cccc}
       C_{12}  &S_{12}&0&0 \\
      - S_{12} &  C_{12} &  0& 0 \\
      0 & 0 &1& 0\\
          0& 0& 0&1
      \nonumber\end{array}\right),\\
    \end{equation}
\begin{equation}\label{M61}
R_{14}=\left(\begin{array}{cccc}
       C_{\alpha} &0 &0&S_{\alpha} \\
        0 & 1 &  0& 0 \\
        0 & 0 &1& 0\\
         -S_{\alpha}& 0& 0&C_{\alpha}
      \end{array}\right),
      R_{24}=\left(\begin{array}{cccc}
      1 &0 &0&0 \\
        0 & C_{\beta} &  0& S_{\beta} \\
       0 & 0 &1& 0\\
          0& -S_{\beta}& 0&C_{\beta}
      \end{array}\right),
 R_{34}=\left(\begin{array}{cccc}
       1  &0&0&0 \\
      0 &  1 &  0& 0 \\
      0 & 0 &C_{\gamma}&S_{\gamma}\\
          0& 0&-S_{\gamma}&C_{\gamma}
      \end{array}\right),\\
    \end{equation}

    \begin{equation}\label{M62}
\phi=\left(\begin{array}{cccc}
       1  &0&0&0 \\
      0 &  e^{i\delta'} &  0& 0 \\
      0 & 0 &1&0\\
          0& 0&0&1
      \end{array}\right),\\
    \end{equation}


\end{document}